\documentclass[%
 aip,
 jmp,%
 amsmath,amssymb,
 reprint,%
]{revtex4-2}

\usepackage{graphicx}
\usepackage{dcolumn}
\usepackage{bm}

\usepackage{moreverb,url}

\usepackage[colorlinks,bookmarksopen,bookmarksnumbered,citecolor=red,urlcolor=red]{hyperref}

\usepackage{fancyhdr}
\usepackage{amsmath}
\usepackage{amsfonts}
\usepackage{amssymb}
\usepackage[makeroom]{cancel}
\usepackage{empheq}
\usepackage{slashed}
\usepackage{epsfig}
\usepackage{makeidx}

\begin{document}


\title[Evolution of timekeeping..., Edval J. P. Santos]{Evolution of timekeeping from water clock to quartz clock - the curious case of the Bulova {\itshape{ACCUTRON 214}} the first transistorized wristwatch}


\author{Edval~J.~P.~Santos}
 \affiliation{Laboratory for Devices and Nanostructures, Engineering at the Nanoscale Group, UFPE, Recife, Brasil.}
 \email{edval@ee.ufpe.br or e.santos@expressmail.dk.}

 \homepage{http://www.ee.ufpe.br.}

\date{\today}

\begin{abstract}
The technological  discoveries and developments since dawn of civilization that resulted in the modern wristwatch are linked to the evolution of Science itself.  A history of over 6000 years filled with amazing technical prowess since the emergence of the first cities in Mesopotamia established by the \v{S}umer civilization.  Usage of gears for timekeeping has its origin in the Islamic Golden Age about 1000 years ago.   Although gears have been known for over 2000 years such as found in the Antikythera Mechanism.  Only in the seventeenth century springs started to be used in clock making.  In the eighteenth century the amazing \textit{Tourbillon} was designed and built to increase clock accuracy.   In the nineteenth century the tuning fork was used for the first time as timebase. Wristwatches started to become popular in the beginning of the twentieth century.  Later in the second half of the twentieth century the first electronic wristwatch was designed and produced, which brings us to the curious case of the Bulova \textit{ACCUTRON} caliber 214 the first transistorized wristwatch, another marvel of technological innovation and craftsmanship whose operation is frequently misunderstood.  In this paper the historical evolution of timekeeping is presented.  The goal is to show the early connection between Science and Engineering in the development of timekeeping devices.  This linked development only became common along the twentieth century and beyond. 
\end{abstract}

\maketitle

\begin{quotation}
  
\textit{This paper is dedicated to the 75 years of the invention of the transistor and to the 25 years of the Laboratory for Devices and Nanosctructures.}

\end{quotation}

\section{Introduction}

Perhaps some 6000 years ago somebody concluded that by using whole fingers alone was not enough for daily counting needs and decided to use knuckles or phalanges too.  To use the hand to count is much better than carry stones.  Thus, using the thumb as indexer one can count to $12$ using the knuckles in the remaining $4$ fingers.    By using both hands one gets $12 + 12$.   But can go further by counting to $12$ with one hand and using the other hand fingers as indexers.  Thus, this procedure can be repeated $5$ times to count up to $60$.  This is a large enough number for most needs.  This could be the origin of base-$60$  number system in the Sumerian civilization.   A truly amazing achievement, \cite{Mella2001,Friberg2008,Friberg2019}.

Using the developed base-$60$ system the Sumerian divided the day in two periods of $12$ hours, created the hour of $60$ minutes, and the minute of $60$ seconds.  The circle was divided into $360$ degrees.  This is approximately the number of days of Earth complete orbit around the Sun.  The year can be divided into $12$ full moons or months.  Thus, $12$ became a magic number used by many later on.   Our lives are synchronized by watches and clocks. Thus, such developments are routinely used in modern day timekeeping but its origin is unknown by most people, \cite{Mella2001,Friberg2008,Friberg2019}.

The goal of this paper is to show the early connection between Science and Engineering in the development of timekeeping devices.   This linked development only became common along the twentieth century and beyond.   The operating principle of the mechanical watch, the {\itshape ACCUTRON} caliber 214 wristwatch, and the modern quartz watch are discussed to demonstrate how the combination of scientific and technological concepts in elasticity, electronics, non-linear feedback control and microelectromechanical systems are linked in development of clocks.   The paper is divided into seven parts, this introduction is the first.  Next, the historical evolution of timekeeping is presented.  The third section reviews elasticity theory as used in clock technology.  The fourth section is about the invention and popularization of the wristwatch.  The fifth section details the operation of the Bulova {\itshape ACCUTRON} caliber 214 the first transistorized wristwatch.  Next the quartz wristwatch is presented, and finally the conclusion.

\section{Early History}

About 10,000 years ago hunter-gatherers domesticated animals and plants and became farmers.  To keep crops they had to live in villages which later became the first cities.  To be able to tell time is a basic skill to live in urban society as synchronization is needed.  Thus, in the dawn of civilization different technologies were developed to tell time.  This includes water clock,  Sun shadow during the day or  burning candle during the night.  About $2200$ years ago the regulated water clock was invented by Ktesibius of Alexandria (285-222bc) and his technology became known as Clepsydra (\textit{water thief}).  This technique improves the accuracy of the clock and is the first known example of equipment with feedback control system, \cite{Lee2008,Lewis1992}.  The usage of gears in clocks was introduced during the Islamic Golden Age in the $1000$ years period from the fall of Rome to the fall of Constantinople.  A wonderful example of clock with gears is the al-Jazari (1136-1206) elephant clock, \cite{al-Jazari1974,Sen2016,RTE2016}.   Although gears have been known for over 2000 years as can be seen in the Antikythera mechanism~\cite{FBM2006,FHDMG2021,Marchant2010}, whose complexity suggests that gears technology may be even older and could be known in Anatolia and beyond.

\begin{figure*}[h]
  \epsfxsize=0.7\linewidth
        \centerline{\epsffile{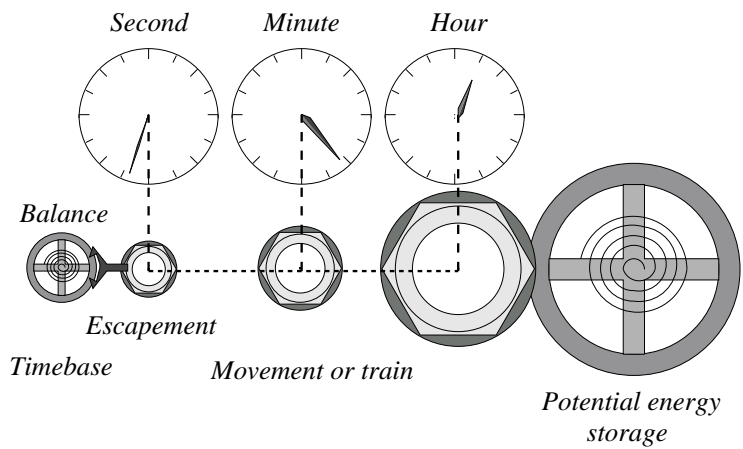}}
\caption{Basic components of the classical mechanical clock.}
\label{fig:mechanicalclock}
\end{figure*}

Chinese developed the base-$10$ system over 3000 years ago~\cite{Temple1988,Temple1986,Needham1956,Katz2009}. This system with the Hindu concept of abstract numbers including zero and Arabic symbols, known as Hindu-Arabic numerals, arrived in Europe translated from ibn Musa al-Khwarizmi (780-850) book named Compendious Book on Calculation by Completion and Balancing (\textit{Al-Kitab al-mukhtasar fi hisab al-jabr wa'l-muqabala}) which became known as Algebra (\textit{al-jabr}), \cite{Katz2009}.  Only after the XVth century this system replaced the Roman numerals in Europe. In the tenth century,   ibn al-Haytham (965-1040) or Alhazen in Latin form introduced the scientific method.  He taught that a hypotheses had to be presented and tested,  diverging from Aristotle (384-322bc) teachings, \cite{TA2007,O1979}.  Geometrical optics, optical lens manufacturing technology, Hindu-Arabic numerals and the scientific method are some of the Arabic knowledge used by Galileo de Galilei (1564-1642).  Galileo improved the scientific method of Alhazen by stating that alternative hypothesis or explanations need to be explored and he used Arabic Mathematics to discover laws of Nature.  This flow of knowledge is the basis of European Renaissance and later the Industrial Revolution, \cite{Lewis1992}.

One day, around 1588, Galileo de Galilei a devoted Catholic was perhaps praying or bored and the movement of a chandelier hanging from the church ceiling called his attention.  It was swinging back and forth.  Later, after 1602, he decided to carry out a careful investigation of the observed pendular motion of the suspended lamp.  He concluded the swinging time period is constant, thus he discovered isochronism of the pendulum.  Galileo de Galilei has acquired scientific knowledge, \cite{Buyse2017}. 

Nowadays, by applying Isaac Newton's (1643-1727) law such observation result can be described mathematically by modeling the suspended lamp as a point ball of mass $m$ attached to a massless one-dimensional rope.  The model point ball exerts a force $F_T$ on the rope of length $L$.  Assuming the oscillation amplitude is sufficiently small,  $F_T\approx mg$,  

\begin{equation}
a=  {d^2x\over dt^2}= -\left({F_T\over m}\right)\mbox{sen}\theta\approx -\left({F_T\over m}\right){x\over L}= -\omega^2 x.
\end{equation}
in which $\omega^2= \left({F_T\over m}\right){1\over L}$.

By definition $ \omega= {2\pi/ T}$.  Thus,

\begin{equation}
\boxed{
    T= 2\pi\sqrt{L\over g}.
}
\end{equation}

The period of oscillation does not depend on the amplitude, as expected for any linear oscillator.  The pendulum movement is isochronous.  A marvelous mathematical prediction!

In 1600 friar Giordano Bruno (1548-1600) was sentenced to death by burning for defending the heliocentric system of Nicolaus Copernicus (1473-1543).    Galileo de Galilei was also condemned for defending this model.  But with the telescope he created, Galileo de Galilei could prove this hypotheses.  As the Hole Office of the Inquisition was haunting the European south, Galileo de Galilei ideas were understood and accepted in north of Europe.   Inhabitants of the north were less sympathetic to the Catholic Church rulings.  In the Netherlands Christiaan Huygens (1629-1695) got interested by Galileo de Galilei results and built the pendulum clock in 1656, named {\em Horologium Oscillatorium}, \cite{Buyse2017}.

In 1672 Christiaan Huygens used a spring to move a balance in a periodic fashion as timebase.  Thus elasticity entered into the construction of the clock to keep time, \cite{Buyse2017}.  The main components of the classical mechanical clock shown in Figure~\ref{fig:mechanicalclock} are:

\begin{itemize}
\item Potential energy source: pendulum, weight, main spring.
\item Movement or train: wheels and pinions of various diameters for energy transfer and timekeeping.  Diameters are selected to achieve the correct rotation for hour, minute, second.
\item Escapement: to avoid the potential energy escaping at once.  Thus, regulating the flow of energy. 
\begin{itemize}
\item Controller mechanism: lever, pallet or index to regulate the speed of the escapement as defined by the timebase. 
\item Timebase or balance:  pendulum, spring or tuning fork . 
\end{itemize}
\item Indicator:  dial for time display with clock hands attached to wheels for hour, minute, and second .
\item Complications:  chronograph, day of week, date, alarms, Zodiac sign, etc.
\item Synthetic crystals, named jewels for marketing purposes,  are used at various positions to reduce friction.
\end{itemize}

In 1795 Abraham-Louis Breguet (1747-1823) invented the {\em Tourbillon} (whirlwind), a mechanism to reduce the negative impact of gravitational force on accuracy of spring-based clock.  The timebase or balance and escapement as shown in Figure~\ref{fig:mechanicalclock} are mounted in a rotating cage.  This is the single axis {\em Tourbillon}.  The mechanism itself is an example of precision craftsmanship and its operation is of extreme beauty.  Specially when manufactured for wristwatches which started to become popular in the beginning of the twentieth century.  For its correct operation the gravitational field must be perpendicular to the {\em Tourbillon} axis, which is easily achieved for desk, tower, even trousers pocket clocks and watches.  But not in a wristwatch.  For wristwatches one can use the triple axis {\em Tourbillon}.  As it is a beautiful mechanical machine, watches with the {\em Tourbillon} can be found even today, \cite{Denny2010}.    In 1866 N. Niaudet and Louis C. Breguet (1804-1883) proposed to replace the spring balance with the tuning fork, \cite{Becket1903}.  Tuning fork based watches became common in the second half of the twentieth century and is the basis of present day electronic watch. 

As a result of many wars in Europe, many fled to the United States of America taking knowledge with them.  This flow of knowledge helped to transform the USA into a world economic power in the second half of the twentieth century.  Similarly, to the flow of knowledge which benefited Europe after the fall of Constantinople.

\section{Linear elasticity}
 
\subsection{Robert Hooke experiment}
The pendulum clock was created Christiaan Huygens in 1656 and he used the spring balance in the construction of a clock in 1672.  Possibly motivated by the technological development in Netherlands,  Robert Hooke (1635-1703) started an investigation  to understand spring behavior circa 1675.  It was necessary to prove the spring balance is isochronous for its application in clocks.    Previously Robert Hooke had worked with pendulum clock and possibly developed the anchor escapement in 1657, \cite{Becket1903}.  However, it seems the anchor escapement was probably developed by William Clement (1638-1704) who was a clockmaker.

It is an observable fact that a solid submitted to mechanical stress deforms or becomes strained.  Robert Hooke concluded that such strain is approximately proportional to the applied stress.   In 1676 he wrote his finding as an anagram in Latin, the language of Science at the time.  He wanted to make sure that he got the credit for the discovery but he decided to keep as a (technological) secret.  In 1678  he disclosed the secret,

\begin{equation}
  \mbox{\centerline{\bf ceiiinosssttuv $\Longrightarrow$  ut tensio, sic vis}}\nonumber
\end{equation}

\noindent or: 
{\centerline  {  ut tensio, sic extensio $\Longrightarrow$ as stress, so strain}}

\begin{equation}
  \boxed{
    F= \kappa\,u
  }
\end{equation}

This is known as Hooke's law or linear elasticity, \cite{Landau1970}.    Robert Hooke and Thomas Tompion (1639-1713) collaborated to build a clock with spring balance similar to the clock independently developed by Christiaan Huygens.  This clock was finished in 1675.

\subsection{Microscopic theory}

To build the mathematical theory of elasticity the solid is divided into parts much smaller than the solid itself, but much larger than the atomic spacing.   Thus, the solid is considered as a continuous medium and the theory of continuous functions can be applied.    The foundations of elasticity theory was laid down by Augustin-Louis Cauchy (1789-1857) and Sim\'{e}on Denis Poisson (1781-1840) in early nineteenth century, \cite{Katz2009,Landau1970}.

\subsubsection{Strain tensor - }

According to Ren\'{e} Descartes (1596-1650), a reference system must be selected so numbers can be assigned to points in the solid for mathematical theory be used to describe physical phenomena.  Consider two arbitrary points separated by a distance vector $\vec{u}= \vec{r'} - \vec{r}$.  After the solid is subjected to deformation, \cite{Landau1970},

\begin{equation}
d\vec{r'}\cdot d\vec{r'}=    d\vec{r}\cdot d\vec{r} + 2\sum_idx_idu_{i} + \sum_i(du_{i})^2.
\end{equation}

As is assumed the solid is continuous and can be described by continuous functions,  $u_i= u_i(x_1,x_2,x_3)$ and $du_{i}= \sum_{k}{\partial u_i\over \partial x_k}dx_k$.    This yields the strain tensor,

\begin{equation}
d\vec{r'}\cdot d\vec{r'}= d\vec{r}\cdot d\vec{r}  + 2\sum_{i,k}u_{ik} dx_kdx_i
\end{equation}
in which

\begin{equation}
  \boxed{
    u_{ik}\equiv{1\over 2}\left[{\partial u_i\over \partial x_k} + {\partial u_k\over \partial x_i}  + \sum_{l} {\partial u_l\over \partial x_k} {\partial u_l\over \partial x_i}\right].
    }
\end{equation}

If $u_{ik}= 0$ only translation or rotation has occurred.  For $u_{ik}\ne 0$ the solid is deformed.  Clearly it is a symmetric tensor,  $u_{ik}= u_{ki}$, hence it can diagonalized at each point in space, $dx_i'= (1+u_{ii})dx_i$.  Thus, to first order,

\begin{equation}
  dV'\approx  \left(1 + \sum_iu_{ii}\right)dV
\end{equation}

\begin{equation}
{dV' - dV\over dV}= \sum_iu_{ii}= \mbox{Tr}(u)= \nabla\cdot\vec{u}.
\end{equation}

The trace of the strain tensor is a measure of relative volume change.

\subsubsection{Stress tensor - }

\begin{figure}[h]
\centering
\includegraphics[width=0.25\linewidth]{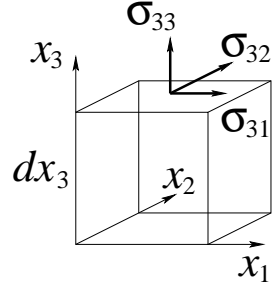}
\caption{Applied stress on infinitesimal cube element of the solid under investigation.}
\label{fig:stress}
\end{figure}

Let us assume the solid is completely neutral, no electric or magnetic fields are created under strain, and gravitational effects are neglected.   Such effects can be dealt with later as a superposition.  Only short range forces are considered. Under such assumptions if the solid is in mechanical equilibrium internal forces are null and two neighbor infinitesimal volume elements interact by the contact surface only, as shown in the infinitesimal cube in Figure~\ref{fig:stress}, \cite{Landau1970}.  

Should an external force be applied, the solid can translate, rotate or deform to reach the new equilibrium condition.  In equilibrium all forces inside the volume element cancel out.    Only forces acting on the surface may not cancel.  Thus  the force per unit volume can be rewritten as the divergence of the force per unit area acting on the external surface of volume element, as stated by the divergence theorem.  Such surface forces is the origin of deformation.

\begin{equation}
 F_i= \int f_i dV= \sum_k\int{\partial \sigma_{ik}\over \partial x_k}dV= \sum_k\oint \sigma_{ik}\hat{n}_kdA_k
\end{equation}
in which

\begin{equation}
  f_i= \sum_k{\partial \sigma_{ik}\over \partial x_k}.
\end{equation}

The components $\sigma_{ik}$ are the force per unit area and form the stress tensor.   Similarly to the force, one assumes that mechanical torque can be written as a surface integral. 

\begin{equation}
   m_{ik}=  \sum_l\oint (\sigma_{il}x_k - \sigma_{kl}x_i)\hat{n}_ldA_l + \int (\sigma_{ki} - \sigma_{ik})dV
\end{equation}

As a consequence the stress tensor has to be symmetric, $\sigma_{ki}= \sigma_{ik}$, \cite{Landau1970}.

\subsection{Microscopic Hooke's law and dynamics}

To obtain the relationship between stress and strain one can apply William R. Hamilton (1805-1865) principle of minimum action, \cite{GPS2001}.  Thus it is required to write down the energy.  For a solid under stress the tensor product between the stress tensor and the strain tensor is the energy density per unit volume,

\begin{equation}
{\cal E}= \int  {\cal U}dV= {1\over 2} \sum_{ij}\int \sigma_{ij}u_{ij}dV.
\end{equation}

Thus it is convenient to use the solid internal energy density and apply Thermodynamics theory.  This theory is a major scientific development of nineteenth century physics.  It displays  extreme beauty as it does not assume internal structure and tries to understand the object or medium under investigation through its interactions, its energy exchanges, with the external environment, \cite{Anderson2005}.

Firstly, the internal energy density, ${\cal U}$, is expanded as polynomial series or Taylor series of the strain tensor.  For a solid in equilibrium,  ${\partial {\cal U}\over \partial u_{ij}}= \sigma_{ij}= 0$.   Defining the stiffness tensor, $c$, the lowest order approximation is quadratic,

\begin{equation}
  {\cal U}= {\cal U}_0 + {1\over 2!}\sum_{ijkl}c_{ijkl}u_{ij}u_{kl} + {\cal O}(2).
  \label{eq:freeenergytaylor}
\end{equation}

Terms such as $ {1\over 3!}\sum_{ijklmn}c_{ijklmn}u_{ij}u_{kl}u_{mn}$ and higher order are related to non-linear elasticity.  According to Thermodynamics theory the internal energy density of a medium at temperature, $T$, with entropy density, ${\cal S}$, under stress,  $\sigma$, satisfies

\begin{equation}
  d{\cal U}= Td{\cal S} - d{\cal W}= Td{\cal S} + \sum_{kl}\sigma_{kl}du_{kl}.
\end{equation}

Experimentally is difficult if not impossible to keep entropy constant as energy is exchanged.  Thus for practical purposes is more convenient to use Helmholtz free energy, ${\cal F}_H= {\cal U} - T{\cal S}$.  Mathematically one introduces a Lagrange multiplier.  Variation of entropy, $d{\cal S}$, is replaced with variation of temperature, $dT$, \cite{Landau1970},

\begin{equation}
  d{\cal F}_H= - {\cal S} dT + \sum_{kl}\sigma_{kl}d u_{kl}.
\end{equation}

Next the principle of minimum action is applied at constant temperature to obtain the relationship between stress, $\sigma_{ij}$, and strain, $ u_{kl}$.  Tensor is an abstract mathematical concept, should a coordinate system be selected, one can represent the tensor by its set of elements. Differentiating Equation~\ref{eq:freeenergytaylor}, one can write

\begin{equation}
  \boxed{
    \sigma_{ij}= \left.{\partial{\cal F}_H\over \partial u_{ij}}\right|_{T}= \sum_{kl}c_{ijkl} u_{kl}
  }
  \label{eq:stiffnesstensor}
\end{equation}
in which $c_{ijkl}$ is the element of the stiffness tensor.

The inverse relation yields the compliance tensor, $s$.

\begin{equation}
  \boxed{
    u_{ij}=  \sum_{kl}s_{ijkl}\sigma_{kl}
  }
  \label{eq:compliancetensor}
\end{equation}

For a dynamical system according to Newton second law,

\begin{equation}
  \sum_j{\partial \sigma_{ij}\over \partial x_j} + f_{ext,i}= \rho  {\partial^2 u_{i}\over \partial t^2}.
\end{equation}

Replacing Equation~\ref{eq:stiffnesstensor} one gets the wave propagation equation, i.e., an equation whose solution can display oscillatory behavior.

\begin{equation}
\sum_{jkl}  {\partial\over \partial x_j}\left(c_{ijkl}{\partial u_k\over \partial x_l}  \right) + f_{ext,i}= \rho  {\partial^2 u_{i}\over \partial t^2}
\end{equation}

Neglecting external force densities, $f_{ext,i}= 0$, and considering plane waves, $u_i= U_i\,e^{j|\vec{k}|(\hat{n}\cdot\vec{x} - v_at)}$, yields

\begin{equation}
\left(\sum_{k} \Gamma_{ik} - \rho v_a^2\delta_{ik}\right)U_k=0
\end{equation}
in which $\Gamma_{ik}= \sum_{jl} c_{ijkl} n_jn_l$ is the Christoffel tensor.

The three roots of the characteristic equation, also known as secular polynomial,

\begin{equation}
  \det\left|\Gamma  - \rho v_a^2\mbox{\bf I}\right|= 0,
\end{equation}
are the acoustic phase velocities: one compression wave and two shear waves.

For a quartz bar along the $y$-axis vibrating in the $x$-direction and energy flow in the $z$ direction, $\hat{n}=(0,0,1)$,

\begin{equation}
\Gamma_{ik}= c_{i3k3}.
\end{equation}

Applying Woldemar Voigt notation, two indexes can be combined into one.

\begin{equation}
  \begin{array}{cc}
    11 &\rightarrow 1\\
    22 &\rightarrow 2\\
    33 &\rightarrow 3\\
    23 &\rightarrow 4\\
    13 &\rightarrow 5\\
    12 &\rightarrow 6\\
  \end{array}
  \label{eq:voigt}
\end{equation}

Thus, the elastic tensor components of quartz can be rewritten as $c_{1313}= c_{55}$,  $c_{2323}= c_{44}$,  $c_{3333}= c_{33}$.  Due to quartz crystal symmetries, $c_{55}= c_{44}$, one gets the following characteristic equation,

\begin{equation}
\left(c_{44}   - \rho v_a^2\right)^2\left(c_{33}   - \rho v_a^2\right)= 0.
\end{equation}

As $f\lambda= v_a$, the shear wave frequency can be determined and it only depends on the microscopic properties of the material,

\begin{equation}
  f= {1\over \lambda}\sqrt{c_{44}\over \rho}.
  \label{eq:shearfreq}
\end{equation}

Next, one has to take energy loss into account.  The  pendulum and spring balance oscillation will fade away with time and eventually stops.  This effect can be included in the mathematical model with the introduction of a dissipative or viscosity term,

\begin{equation}
\sum_{jkl}  {\partial\over \partial x_j}\left(c_{ijkl}{\partial u_k\over \partial x_l}  \right)= \rho  {\partial^2 u_{i}\over \partial t^2} + \nu_f   {\partial u_{i}\over \partial t}
\end{equation}
in which $\nu_f$ is the viscosity coefficient.

To keep oscillation is required a mechanism to replenish the lost energy.  The Elgin, LIP and Hamilton Electric 500 wristwatches realizes this mechanism with the tiny electromagnetic hammer, \cite{Alft2003,Rondeau2006}.  There is no feedback.  In the Bulova ACCUTRON wristwatch the transistor implements a feedback mechanism, a concept created by Ktesibius, \cite{Lee2008,Bulova1960}.  The amount of energy loss is detected and the electronic circuit replenish to keep the oscillation running with constant amplitude as long as the battery is charged.  This is discussed next.

\section{Birth of the electronic wristwatch}

In 1905 Alberto Santos-Dumont (1873-1932) asked his friend Louis Cartier (1847-1942) to design a watch which could be held around the wrist because he had difficulty to reach the watch in the pocket during his flights.  This is the first wristwatch to become popular among men.  Many variants of Santos watch are still manufactured by Cartier, \cite{Cologni2017,CM1993}.  Before this a few wristwatches were made as a piece of jewelry for noblewomen.  One was designed  by Abraham-Louis Breguet in 1810 and another by Antoni Patek and Adrien Philippe in 1868.

\begin{figure}[h]
\centering
\epsfxsize=0.7\linewidth
\centerline{\epsffile{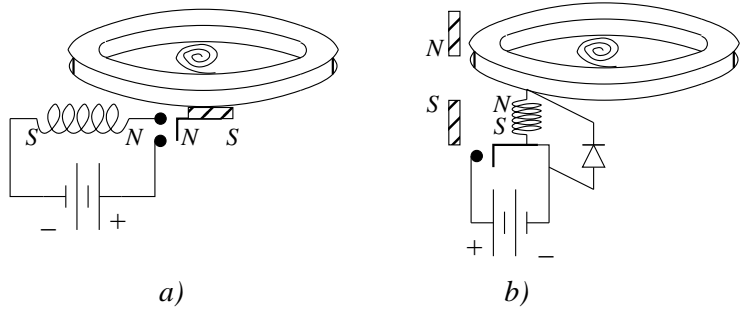}}
\caption{a) Simplified electric circuit  of the Elgin and LIP wristwatch.  b)   Simplified electric circuit  of the Hamilton 500 wristwatch.}
\label{fig:elginhamilton}
\end{figure}

Electric clocks were developed and patented in 1841 by clock maker Alexander Bain (1810-1877).  He used in his design the electromagnet technology developed a couple of decades earlier.  In 1952 Elgin Watch Company announced the first electric wristwatch which uses a spring balance, an electric coil (electromagnet) and a permanent magnet, as shown in Figure~\ref{fig:elginhamilton}a.

At each spring balance swing, electric contact is made and a small repulsive force is generate to restore the spring movement.  The electric system acts as a tiny hammer.  This will keep the movement while the battery lasts.  Eliminating the need to wind the watch.  The concept is very much identical to the electric clock of Alexander Bain.  

The hammer force generated to boost the spring balance movement is directly proportional to the product of the permanent magnet dipole, $m_1= {B_R\over \mu_o}V_1$, and the electromagnetic dipole, $m_2$, created by the electric current as the coil makes contact at each cycle.  Assuming an electric coil with radius much smaller than its length, the coil magnetic moment is $m_2= N_2I_2A_2$.  Both dipoles are aligned in the same axis,  separated by a distance $r$, and oriented in the opposite direction, so the generated force is repulsive.  The approximate expression for the hammer force,

\begin{eqnarray}
  F(r,m_1,m_2)=&& {3\mu_o\over 2\pi r^4}m_1m_2\\
 =&& {3\mu_o\over 2\pi r^4}\left({B_R\over \mu_o}V_1\right)\left(N_2I_2A_2\right)\\
 =&& {\cal K}I_2.
\end{eqnarray}

Elgin Watch (USA) collaborated with LIP Watch (France) to develop a commercial product but took longer than expected.  In 1957 the Hamilton Electric 500 wristwatch was released with the ideas presented by Elgin Watch, \cite{Alft2003,Rondeau2006}.  The main difference is that in the Hamilton Electric 500 the magnet was fixed and the coil was attached to the spring balance, as presented in Figure~\ref{fig:elginhamilton}b.

Hamilton Electric 500 displays an extra feature: a diode is added as protection device.   As the electric coil is switched on and off a high voltage appears at its terminals, $V= L(d I/ d t)$, which generates sparks.  A diode provides a path to the electric current dissipating the stored magnetic energy.  Due to this electronic protection device, the Hamilton Electric 500 is known as the first electronic watch.  But they are better classified as electric wristwatches.  In any case both watches switch the electric circuit to save energy as electric button batteries appropriate for watch use had little electric charge capacity and was very expensive.

\section{Bulova ACCUTRON 214 caliber wristwatch}

Late nineteenth century Joseph Bulova (1851-1936) decided to immigrate from Check Republic to USA.  In 1875 he founded Bulova Company in New York for production of watches.  After the release of the electric watch, Bulova Watch Co. was afraid the new electric watch could affect its business harshly.   It was decided to develop its own electric wristwatch.  Engineer and inventor Max Hetzel (1921-2004), a Bulova employee, was assigned with the task to evaluate the new electric watches and start the development of a new wristwatch for Bulova.  He concluded that the electric wristwatches already available using spring balance technology did not improve accuracy.    A more accurate watch could be the key to gain market share.  Thus, he designed the first wristwatch with the newly developed transistor and the tuning fork for timekeeping as proposed by Louis Breguet over $50$ years earlier.   The Bulova wristwatch with commercial name \textit{ACCUTRON}, \textit{ACCUracy through elecTRONics}, was released in 1960, \cite{Lee2008,Sigmond2018,Bulova1960}. 

The Bulova \textit{ACCUTRON} caliber 214 wristwatch can be considered the first real electronic wristwatch.  The timebase is a tuning fork oscillating at  $360\,$Hz.  At this frequency the tuning fork produces an audible humming sound, an untuned F4 ($349.228$\,Hz) musical note instead of the traditional tic-tac, which characterizes the mechanical wristwatch.   This higher pitch is a key design feature to improve accuracy.  By dividing the higher frequency timebase down to $1$\,Hz, accuracy is improved.  The Bulova \textit{ACCUTRON} caliber 214 wristwatch accuracy is  $2\,$seconds a day or $1\,$minute a month.   This represents a great improvement in the art of timekeeping as compared to all competing models.

\subsection{Operating principle}

\begin{figure}[h]
        \epsfxsize=0.7\linewidth
        \centerline{\epsffile{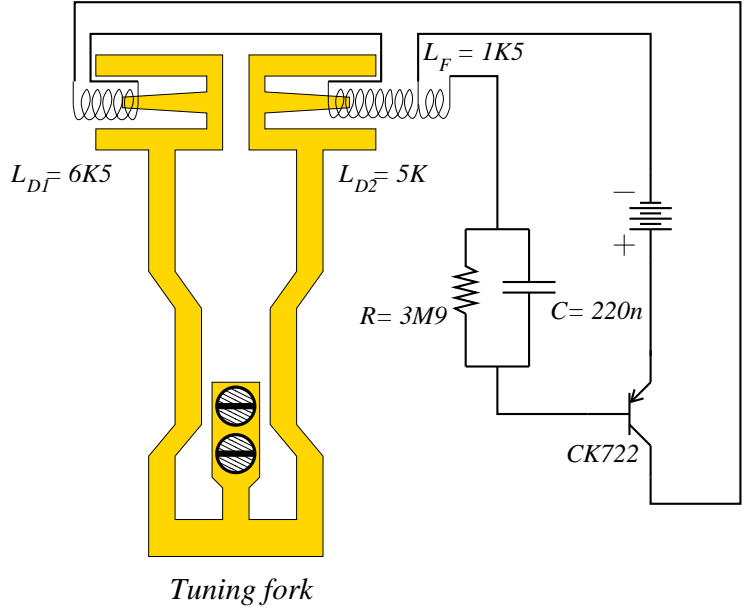}}
\caption{Electromechanical circuit of the Bulova \textit{ACCUTRON} caliber 214 wristwatch with the CK722 germanium PNP transistor.}
\label{fig:bulovaaccutron}
\end{figure}

The electromechanical circuit of the Bulova \textit{ACCUTRON} caliber 214 wristwatch is shown in Figure~\ref{fig:bulovaaccutron}.   There is no switching as in  previously released electric wristwatches.   Each coil is made of copper ($\rho= 1.68\times 10^{-8}\Omega$.m) with  $16\,\mu$m diameter and $80$ meters long wire.  By itself a demonstration of high technical skill.   For such specification the coil resistance is $R_{coil}\approx 6685\,\Omega$.  

As the permanent magnet attached at the tuning fork tine ends swings back and forth inside the coil, a periodical voltage signal is induced at the coil terminals, which can be approximated by its fundamental component, 

\begin{equation}
V_L= {d\over dt}\int B_M\cdot\hat{n}dS= V_A\sin{\omega t}.
\label{eq:inducedvoltage}
\end{equation}

The transistor is model CK722 made of germanium and manufactured by Raytheon.  Originally developed for hearing aid equipment and later presented to the hobbyist market to build radio receivers, \cite{Lee2008,PopularMechanics1953}.  The electronic circuit in the Bulova \textit{ACCUTRON} caliber 214 watch is quite similar to the circuit presented as a simple radio receiver for hobbyist, except that in the watch the germanium PNP transistor in the common emitter configuration is part of a non-linear feedback network designed to keep the tuning fork oscillation amplitude constant.   As can be seen in Figure~\ref{fig:bulovaaccutron}, a capacitor, $C=  220$\,nF, and a resistor, $R=  3.9$\,M$\Omega$, are in parallel and attached to the base of the PNP transistor.   At the oscillating frequency the capacitive reactance is much smaller than the resistance,

\begin{eqnarray}
  X_C&=& {1\over \omega C}= {1\over 2\pi 360 \times 220 \times 10^{-9}}=  2\,k\Omega \nonumber\\
          &\ll&  R= 3.9\,M\Omega\nonumber.
\end{eqnarray}

Thus, the capacitor is a short circuit for the AC signal generated by the tuning fork oscillation.  The resistor is used to establish the bias current so the transistor operates in the active region.   For the PNP germanium transistor used $V_{EB}= 0.3\,$V, and $\beta\approx 100$.   The original battery voltage is $1.35\,$V.  Thus, bias current values and small signal amplifier voltage gain can be estimated,

\begin{eqnarray}
  I_B&=& {1.35 - 0.3\over 3.9\times 10^6}= 0.27\,\mu A,\\
  I_C&=& \beta I_B\approx 100 I_B=  26.9\,\mu A,\\
  A_v&=& {I_C\over V_t}X_L\approx   26.
\end{eqnarray}
 
To understand the operating principle one has to follow the feedback loop.  A fraction of the induced signal in Equation~\ref{eq:inducedvoltage} is collected by coil $L_R$ and injected via the capacitor acting as a short-circuit into the transistor base.  This modulates the voltage between the emitter and base which varies the collector current exponentially,

\begin{equation}
I_c= I_S\left(\, e^{V_{eb}\over V_t}-1\right)\approx I_{C}\,e^{v_{eb}\over V_t}
\label{eq:collectorcurrent}
\end{equation}
in which $V_t=k_BT/q$ is the thermal voltage and $V_{eb}= V_{EB} + v_{eb}$.

At small oscillation amplitudes the signal is amplified with the small signal gain.  This increases the collector current injected into the coils which increases the magnetic force in-phase with the tuning fork mechanical oscillation.  This generates a push to increase the amplitude keeping the oscillation going.  

In this design the maximum amplitude is determined by the battery voltage.   The exponential collector current variation in Equation~\ref{eq:collectorcurrent} makes this oscillator non-linear.  As the collector voltage swing approaches the maximum amplitude the differential gain becomes zero stopping any further amplitude increase.  Thus a well defined oscillation amplitude is achieved which is characteristic of non-linear oscillators, i.e., a limit cycle is reached, \cite{Santos2021,Vittoz2010}.

This is a simple but ingenious engineering solution in its most refined sense.  The elegance is in its simplicity.  The drawback is that the amplitude of oscillation is dependent on the battery voltage.  But this could not be avoided with the electronics technology available at the time.     The original battery was mercury based chemistry.  Such batteries are no longer manufactured due to environmental issues.  Should the battery be replaced with modern version at  $1.55\,$V the higher oscillation amplitude and higher collector current can cause premature failure of the coils and tuning fork.  Many variants of this design were made.  The concepts introduced by Max Hetzel in wristwatch design are used today in the quartz wristwatch as discussed next.

\section{Quartz wristwatch}

The fast advancement of electronics technology since the invention of the germanium point-contact transistor in 1947 by William Schokley (1910-1989), John Bardeen (1908-1991) and Walter H. Brattain (1902-1987), integrated electronics in 1949 by Werner Jacobi (1904-1985) and Jack Kilby (1923-2005), and MOSFET fabrication in 1959 by Mohamed M. Atalla (1924-2009) and Dawon Kahng (1931-1992), brought low-voltage low-power integrated circuit technology to the wristwatch in 1967 \cite{Lee2008,Ross1998,Riordan1999,Jacobi1949,Sah1988}.  Watches and clocks with digital electronics technology combined with quartz timebase are much more accurate, consumes less power and are cheaper.  As quartz and integrated circuits entered the wristwatch industry in the 1970's it became known as the \textit{quartz crisis}, \cite{SD2000}.

\begin{figure}[h]
        \epsfxsize=0.7\linewidth
        \centerline{\epsffile{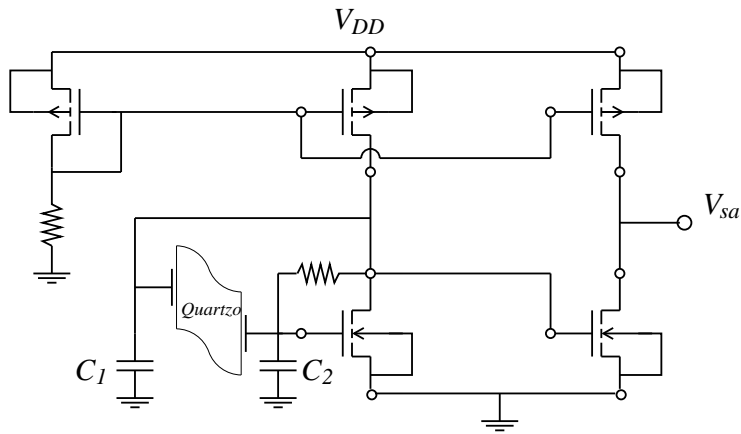}}
\caption{CMOS Pierce oscillator quartz crystal acting a high-Q inductor, $Z_Q= R_Q + j\omega L_Q$.}
\label{fig:pierce}
\end{figure}

Linear oscillators are isochronous at any oscillation amplitude, but do not exist in the real world.  Real oscillators are not linear.  They are designed to have stable limit cycle, which yields stable amplitude and constant period.   Modern wristwatch uses quartz crystal for timekeeping.  It can be used as part of a circuit such as the oscillator invented by George W. Pierce (1872-1956), as shown in Figure~\ref{fig:pierce}.   In this circuit the quartz crystal behaves as a high-Q inductor, $Z_Q= R_Q + j\omega L_Q$ and is part of the non-linear feedback loop of a CMOS amplifier.   Neglecting parasitics, replacing the crystal with the equivalent model and applying electrical circuit theory,  the oscillator behavior can be described as a third order non-linear differential equation with energy loss.  The function  $f(\cdot)$ represents the amplifier circuit used to supply the lost energy, \cite{Santos2021,Vittoz2010},

{
\begin{equation}
{d^3 v_1\over dt^3} + {R_Q\over L_Q}{d^2 v_1\over dt^2} + {C_1+C_2\over L_QC_1C_2} {d v_1\over dt} +  {f(v_1)\over L_QC_1C_2}= 0
\label{eq:thirdorder}
\end{equation}
}

By setting $v= v_1$, $u= dv_1/dt$ and $w= d^2v_1/dt^2$ Equation~\ref{eq:thirdorder} can be rewritten as a set of first order equations in matrix form to generate a phase space plot which is used to locate the limit cycle.  

\begin{eqnarray}
{d\over dt}\left(\!
\begin{array}{c}
v\\
u\\
w\\
\end{array}
\!\right)
&=&
\left(\!
\begin{array}{ccc}
0 & 1 & 0\\
0 & 0 & 1\\
0 &  - {C_1+C_2\over L_QC_1C_2}   &  - {R_Q\over L_Q}\\  
\end{array}
\!\right)\left(\!
\begin{array}{l}
v\\
u\\
w\\
\end{array}
\!\right) - \nonumber\\
&&
\left(\!
\begin{array}{c}
0\\
0\\
{f(v)\over L_QC_1C_2}\\
\end{array}
\!\right)
\end{eqnarray}


The trace of the matrix is the loss coefficient and the determinant is the square of the oscillation angular frequency.  The condition for onset of oscillations is the minimum energy to cancel dissipation.

\begin{equation}
{d^2 v_1\over dt^2} +  {1\over R_QC_1C_2}f(v_1)= 0
\end{equation}

The initial oscillation is a small signal, thus the linear approximation holds, $f(v_1)= g_mv_1$,

\begin{equation}
I_{D,min}=\omega^2R_QC_1C_2(V_{GS}-V_T)
\end{equation}
in which $V_T$ is the MOSFET threshold voltage.

In designing the quartz based oscillator one has to consider other factors such as: parasitics ($\sim 0.5\,$pF), manufacture tolerance, temperature stability, frequency shift (pushing/puling), soldering effects, aging process during life cycle.  For the quartz wristwatch the oscillation frequency is set to  $32768\,$Hz to improve accuracy and cannot be detected by the human ear.

\subsubsection{Quartz - } 
\begin{figure*}[h]
        \centerline{\epsfxsize=1.6in \epsffile{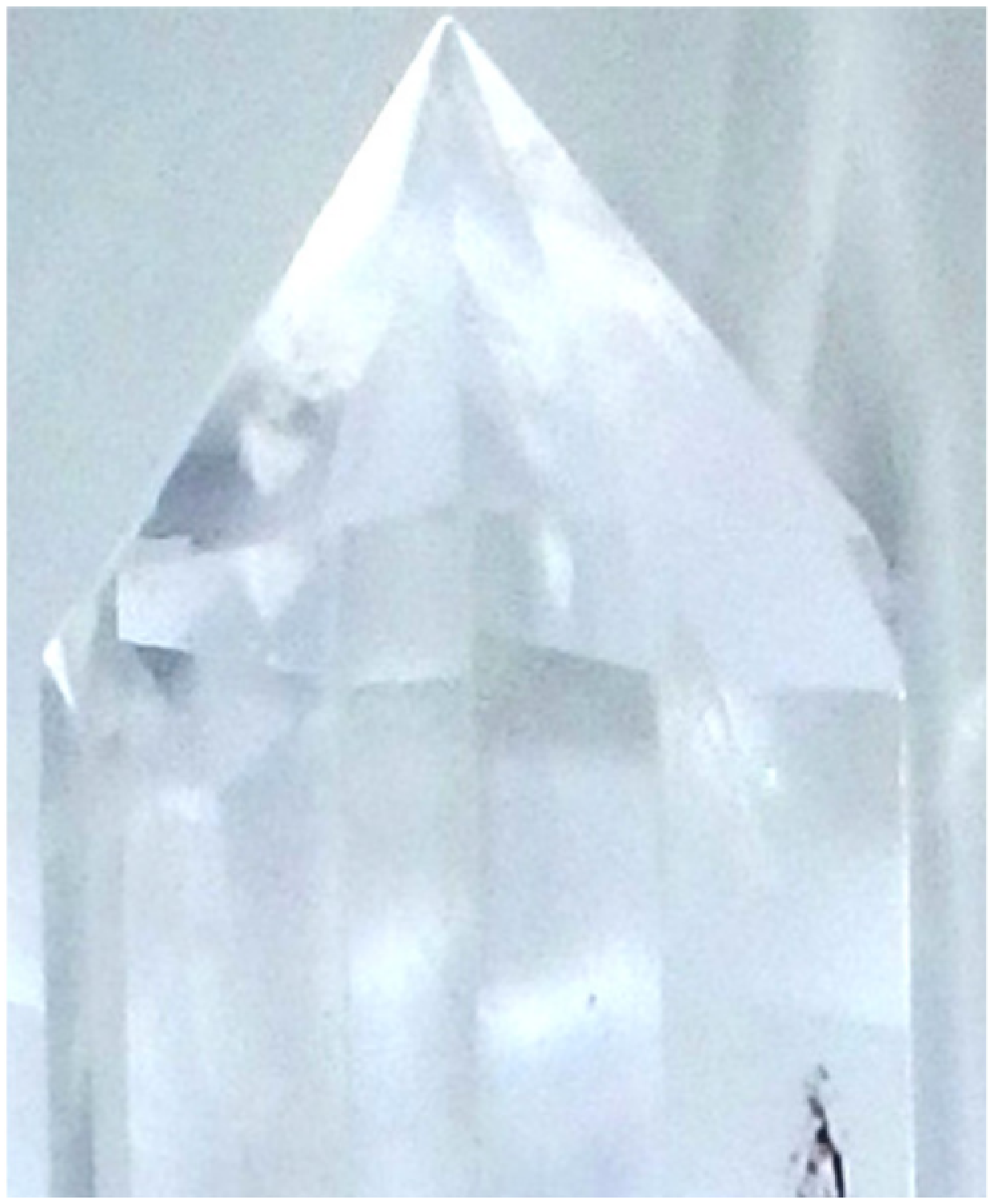}\hskip 0.4in\epsfxsize=1.6in\epsffile{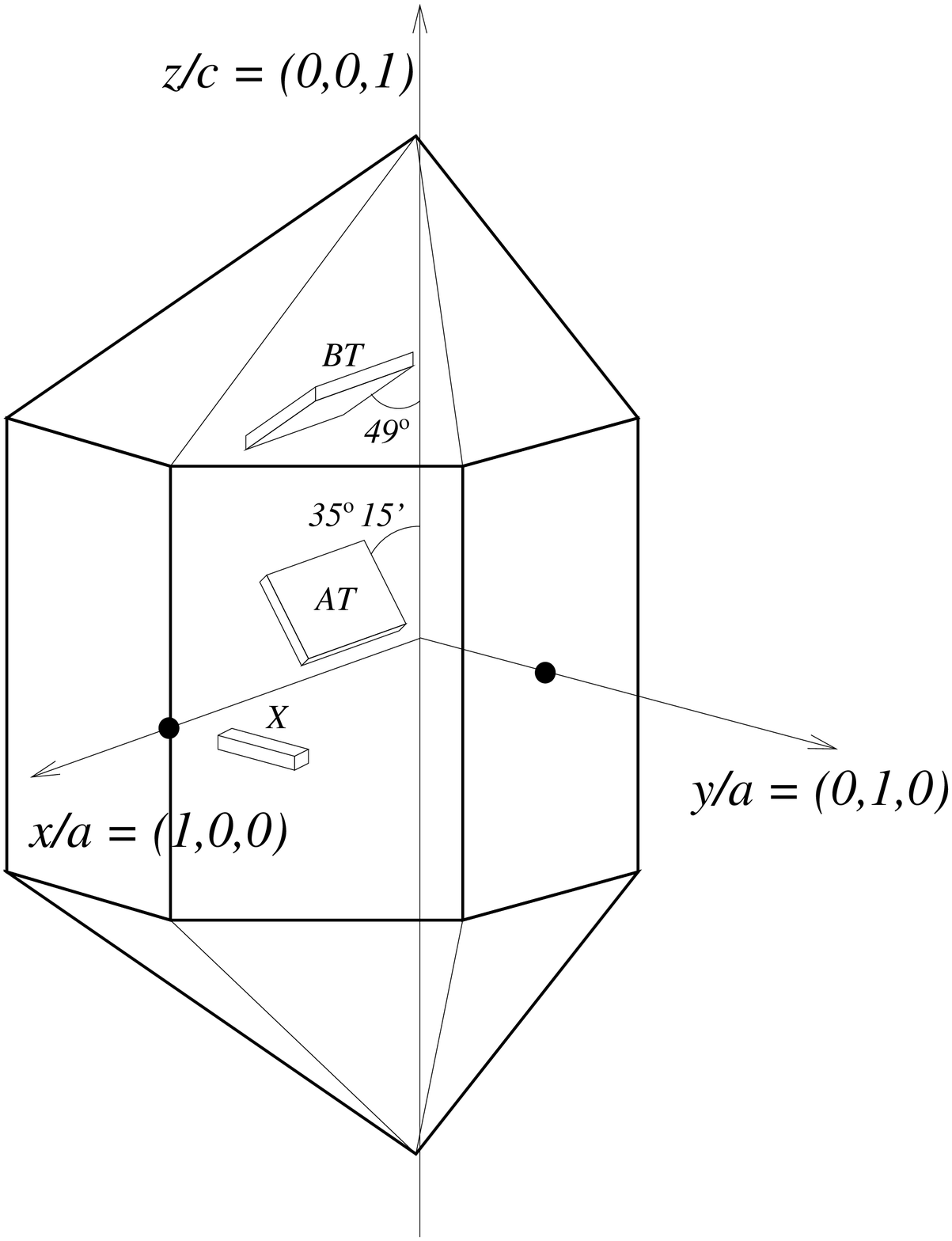}}
\caption{Quartz crystal as found in Nature or synthetically grown, and standard Cartesian axis used to determine the elastic tensor.}
\label{fig:quartz}
\end{figure*}

is the crystalline form of silicon dioxide, SiO$_2$.  Below  $573\,$K  is named $\alpha$-quartz and above this temperature is named $\beta$-quartz.  The crystal cell is classified as trigonal.  Macroscopically it displays an hexagonal symmetry, as shown in Figure~\ref{fig:quartz}, and comes in two flavors: right handed or left handed, \cite{IEEEStd1988,HLK2003}.

Using crystal symmetries and applying Woldemar Voigt notation in Equation~\ref{eq:voigt}, the elastic tensor or stiffness, $c$, for $\alpha$-quartz in GPa,  based on IEEE 176-1987, \cite{IEEEStd1988}, is given by,

{\small
\begin{equation}
\left(
  \begin{array}{cccccc}
    86.74  &  6.99 & 11.91 & -17.91 &        0 & 0 \\
     6.99  & 86.74 & 11.91 & 17.91  &        0 & 0 \\
    11.91  & 11.91 & 107.2 &     0  &        0 & 0 \\
   -17.91  & 17.91 &  0    & 57.94  &        0 & 0 \\
    0      & 0     &  0    &      0 &    57.94 & -17.91 \\
    0      & 0     &  0    &      0 &    -17.91& 39.88 \\
  \end{array}
  \right)
\end{equation}
}

For the trigonal crystal cell $c_{66}= {c_{11}- c_{12}\over 2}= {86.74 -  6.99\over 2}= 39.875$.  The inverse of elastic matrix, $c$, is the compliance matrix, $s$.  The Young modulus is the inverse of the compliance matrix element, $E_{Y,ii}= s_{ii}^{-1}$.

\subsubsection{Quartz disk plate - } is the simplest geometry to machine for the timebase oscillator.  To connect the plate to the circuit, electrodes are attached on both sides.  The oscillator can be tuned to the plate fundamental mode as it vibrates in the shear mode with wavelength twice the plate thickness, $\lambda= 2a_e$.  Considering the frequency of  $32768\,$Hz, according to Equation~\ref{eq:shearfreq},

\begin{equation}
  a_e=  {\lambda\over 2}= {1\over 2\times 32768}\sqrt{57.94\times 10^9\over 2649}= 71\,\mbox{mm.}
\end{equation}

Not an adequate plate thickness for wristwatch.  Nevertheless, it is used with pressure and temperature sensors, analytical balance, among others.  In such applications a much higher frequency is used which makes the plate much thinner, \cite{SS2019,SS2022}.  For example, $f_{Pierce}= 10\,$MHz,

\begin{equation}
  a_e=   {1\over 2\times 10^7}\sqrt{57.94\times 10^9\over 2649}= 0.234\, \mbox{mm.}
\end{equation}

\subsubsection{Quartz tuning fork - } 
\begin{figure}[h]
        \epsfxsize=0.22\linewidth
        \centerline{\epsffile{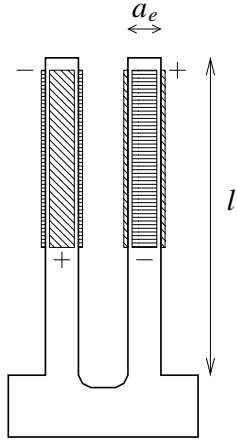}}
\caption{Watch tuning fork is designed to oscillate at $32768\,$Hz.}
\label{fig:diapasao}
\end{figure}

offers a mode of vibration which is capable of vibrating at $32768\,$Hz with much smaller dimensions.   The quartz etching process was developed by J\"{u}rgen Staudt, \cite{Staudte1973}.  The miniaturized turning fork design is shown in Figure~\ref{fig:diapasao}.   This is the same concept as used in the Bulova \textit{ACCUTRON} caliber 214.  The tine (or prong) is a quartz bar along the crystal $y$-axis.  This is known as the X-cut  seen in Figure~\ref{fig:quartz} and it vibrates in the $xy$-plane.   The tuning fork frequency of vibration of the first harmonic, \cite{Landau1970,HBW1999,RRB1992}.

\begin{equation}
  f_{T\!F}= {(1.194\pi)^2 K\over 8\pi l^2}\sqrt{E_Y\over \rho}= {(1.875104)^2a_e\over 2\pi l^2}\sqrt{E_Y\over 12\rho}
\end{equation}
in which $K= a_e/\sqrt{12}$ is the radius of gyration of the rectangular beam cross-section.

For the quartz tuning fork, $E_Y= 78.7\,$GPa,  with tine length,  $l= 2.978\,$mm, and frequency, $f_{T\!F}= 32768\,$Hz the thickness of the tine can be estimated.

\begin{eqnarray}
a_e&=& f_{T\!F}\left[{(1.875104)^2\over 2\pi l^2}\sqrt{78.7
\times 10^9\over 12\times 2649}\right]^{-1}\\
&=&  0.33\,\mbox{mm}
\end{eqnarray}

\begin{figure}[h]
        \epsfxsize=0.6\linewidth
        \centerline{\epsffile{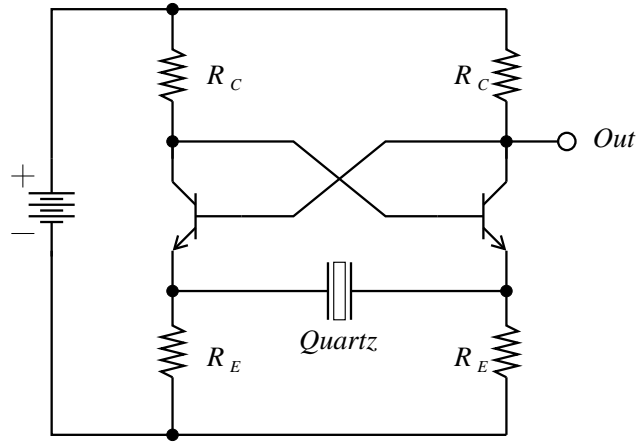}}
\caption{Bipolar oscillator used in the first quartz wristwatch.}
\label{fig:quartzcircuit}
\end{figure}

With such dimensions the quartz tuning fork can be utilized in a wristwatch.  In 1968 the first quartz wristwatch with bipolar transistor and tuning fork was developed independently and presented to the public in Switzerland by Ren\'{e} Le Coultre, Roger Wellinger and Max Forrer, and Japan by Tsuneya Nakamura, \cite{Trueb2010}.  The oscillator circuit is shown in Figure~\ref{fig:quartzcircuit}. 

\begin{figure*}[th]
\centering
\includegraphics[width=1\linewidth]{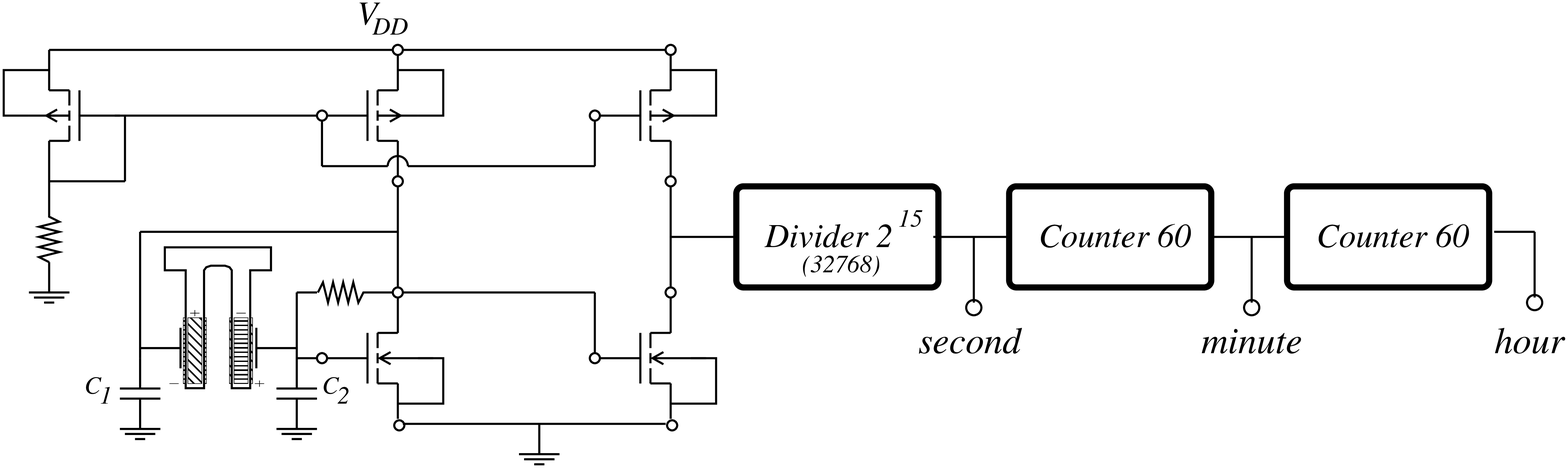}
\caption{CMOS Pierce oscillator with quartz tuning fork vibrating at $32768\,$Hz.  The signal is divided electronically by $2^{15}$ to get the $1\,$s time base.}
\label{fig:piercewatch}
\end{figure*}

Modern quartz wristwatch can use low-voltage low-power CMOS integrated circuit technology.  The Pierce oscillator can be used to build the timebase circuit.  The miniaturized tuning fork can be manufactured with precision micro-electromechanical systems techniques, such as: deep reactive etch, ultrasound machining, wet etch.  The quartz wristwatch frequency of 32768\,Hz is used because this value is a power of $2$,
\begin{equation}
 32768= 2^{15}.  
\end{equation}
Thus, it is a convenient number to carry out division with digital electronics techniques to extract the $1$\,s timebase.   Dividing the oscillator frequency $32768\,$Hz by $2^{15}$ one gets the seconds which are counted to $60$ to get the minutes, which are counted  to $60$ to get the hours, as shown in Figure~\ref{fig:piercewatch}.

\vskip 0.1in
\parbox{5in}{ \bf
Thus, time division scheme developed by \v{S}umer civilization over 6000 years ago is the system that controls the daily life of modern society. }
\vskip 0.1in

The wristwatch selected by NASA for the APOLLO mission was the Omega Speedmaster, a mechanical wristwatch.   Among the specifications  to be met included a temperature range from  $-162^o$C to $126^o$C which is not achieved by the Bulova \textit{ACCUTRON} electronic watch.    Nevertheless, the \textit{ACCUTRON} was taken to the Moon as timekeeping devices in some equipment, including the lunar vehicle left on the Sea of Tranquility.  In the APOLLO XV flight a privately owned Bulova \textit{ACCUTRON} was again taken to the Moon.  This \textit{ACCUTRON} was sold for over 1.6 million dollars.  Another \textit{ACCUTRON} watch was buried in New York city so that future  archaeologists, perhaps 6000 years from now, can appreciate the ingenuity of twentieth century engineering.

\section{Conclusion}

Development of timekeeping instruments over thousands of years is a history of scientific discovery and superb craftsmanship.  Engineering at the highest level of ingenuity.  Scientific and technological knowledge has been created all over the world and its exchange has benefited humanity.   This is oldest example of science and technology being combined which only became common along the twentieth century and beyond.

Technological demand induced scientific discovery which impacted technology.  It is presented how elasticity, piezoelectricity, mechanics, feedback control, materials technology, electric battery technology, analog and digital electronics, and micro-electromechanical technology  are combined to achieve the development of a high precision technological product.  It is shown why the replacement of Bulova \textit{ACCUTRON} original button battery with modern equivalent may lead to reliability problems.  The electronics wristwatch development has an impact in the early development of low-voltage low-power electronics.

After the quartz crisis a renewed interested arose towards mechanical wristwatches as pieces that catches our attention in a ever faster and crazy world and invites us to slow down as we stop to admire the \textit{Tourbillon} doing its job ... or not.

\vskip 0.1in
\parbox{5in}{{\em Science can amuse and fascinate us all,  but it is Engineering that changes the World.}  Isaac Asimov (1920-1992)}

\setcounter{secnumdepth}{0}
\section{Acknowledgements}

This paper is dedicated to the 75 years of the invention of the transistor and to the 25 years of the Laboratory for Devices and Nanosctructures, UFPE (1996-2021).  Most historical dates were fetched initially from Wikipedia and confirmed with other bibliographical references.  

The author states that there is no conflict of interest. 


\bibliographystyle{IEEEtran}
\bibliography{IEEEabrv}

\end{document}